\documentclass[submission,copyright,creativecommons]{eptcs}
\providecommand{\event}{FOCLASA 2015} 
\usepackage{breakurl}             

\usepackage{csquotes}
\usepackage{graphicx}
\usepackage{balance}
\usepackage{longtable}
\usepackage{url}
\usepackage{lscape}

\usepackage{xcolor}
\usepackage{amsmath}
\usepackage{amssymb}
\usepackage{amsxtra}

\usepackage{enumitem}

\usepackage{colortbl,multirow,hhline}

\usepackage{ifthen}
\newboolean{showcomments}
\setboolean{showcomments}{true} 
\ifthenelse{\boolean{showcomments}}
  {\newcommand{\nb}[2]{
    \fcolorbox{gray}{yellow}{\bfseries\sffamily\scriptsize#1}
    {\sf\small$\blacktriangleright$\textit{#2}$\blacktriangleleft$}
   }
   
  }
  {\newcommand{\nb}[2]{}
   
  }

\urldef{\mailsa}\path|{marco.autili,massimo.tivoli}@univaq.it|

\newcommand{\pedice}[1]{\raisebox{-0.6ex}{\scriptsize #1}}

\usepackage{listings}
\usepackage{caption}
\DeclareCaptionFont{white}{\color{white}}
\DeclareCaptionFormat{listing}{\colorbox{gray}{\parbox{\textwidth}{#1#2#3}}}
\definecolor{gray}{rgb}{0.4,0.4,0.4}
\definecolor{darkblue}{rgb}{0.0,0.0,0.6}
\definecolor{cyan}{rgb}{0.0,0.6,0.6}

\lstdefinelanguage{XML}
{
	showstringspaces=false,
	morestring=[b]",
	morestring=[s]{>}{<},
	morecomment=[s]{<?}{?>},
	stringstyle=\color{black},
	identifierstyle=\color{darkblue},
	keywordstyle=\color{cyan},
	morekeywords={xmlns, version, type, encoding, xs:schema,xs:element,xs:complexType,xs:sequence,xs:attribute},
	basicstyle=\ttfamily\scriptsize,
	columns=fullflexible,
	showstringspaces=false,
	commentstyle=\color{gray}\upshape,
	breaklines=true,
	numbers=left,                    
	numbersep=5pt,                   
	numberstyle=\tiny\color{gray},   
	xleftmargin=5pt
}

\title{On the Automated Synthesis of Enterprise Integration Patterns to Adapt Choreography-based Distributed Systems}

\author{Marco Autili \qquad Amleto Di Salle\\ Alexander Perucci \qquad Massimo Tivoli
\institute{Department of Information Engineering Computer Science and Mathematics}
\institute{University of L'Aquila - ITALY}
\email{marco.autili@univaq.it \qquad amleto.disalle@univaq.it}  \email{alexander.perucci@graduate.univaq.it \qquad massimo.tivoli@univaq.it}
}

\def\titlerunning{Choreography Adaptation Through Enterprise Integration Patterns}
\def\authorrunning{M. Autili \& A. Di Salle \& A. Perucci \& M. Tivoli}

\begin{document}
\maketitle

\begin{abstract}

The Future Internet is becoming a reality, providing a large-scale computing environments where a virtually infinite number of available services can be composed so to fit users' needs. Modern service-oriented applications will be more and more often built by reusing and assembling distributed services. A key enabler for this vision is then the ability to automatically compose and dynamically coordinate software services. Service choreographies are an emergent Service Engineering (SE) approach to compose together and coordinate services in a distributed way. When mismatching third-party services are to be composed, obtaining the distributed coordination and adaptation logic required to suitably realize a choreography is a non-trivial and error prone task. Automatic support is then needed. In this direction, this paper leverages previous work on the automatic synthesis of choreography-based systems, and describes our preliminary steps towards exploiting Enterprise Integration Patterns to deal with a form of choreography adaptation.
\end{abstract}

\section{Introduction}\label{sec:introduction}

\vspace{-0.2cm}

The Future Internet promotes a distributed computing environment that will be increasingly surrounded by a large number of software services, which can be composed to meet user needs. 
The Future Internet of Services paradigm emerges from the convergence of the Future Internet (FI) and the Service-Oriented Computing (SOC) paradigm~\cite{FIRE:15}. Services play a central role in this vision as effective means to achieve interoperability between heterogeneous parties of a business process, and
new value added service-based systems can be built as a {\em choreography} of services available in the FI. Service choreography is a decentralized approach, which provides a loose way to design service
composition by specifying the participants (i.e., roles) and the (message-based) interaction protocol among them.
It decouples the participant tasks from the services that only later will be bound to the specified roles.

The need for service choreography was recognized in the Business Process Modeling Notation version 2.0\footnote{\url{http://www.omg.org/spec/BPMN/2.0/}} (BPMN2), which introduced \emph{Choreography Diagrams} to offer choreography-specific modeling constructs. A choreography diagram models peer-to-peer communication by defining a multi-party protocol that,
when put in place by the cooperating parties, will permit to reach the overall choreography objectives in a
fully distributed way. In this sense, service choreographies are quite different from service orchestrations
in which a single stakeholder centrally plans and decides how an objective should be reached through the
cooperation with other services.

In BPMN2 Choreography Diagrams, a role models the expected behaviour (e.g., the expected protocol) that a concrete participant service should match in order to be able to play the role in the choreography. Choreography roles can be specified by using different notations aiming at modeling different views of the expected service behaviour, e.g., WSDL\footnote{\url{http://www.w3.org/TR/wsdl}} for protocol signature specification, BPEL\footnote{\url{https://www.oasis-open.org/committees/download.php/23964/wsbpel-v2.0-primer.htm}} or automata-based notations for protocol interaction specification.

In this paper we leverage the experience on the automatic synthesis of the choreography-based coordination logic for service-oriented systems that we have been doing so far within the EU CHOReOS project\footnote{\url{http://www.choreos.eu/}}. Then, being supported by the EU CHOReVOLUTION (follow-up) project\footnote{\url{http://www.chorevolution.eu/}}, we report on the novel idea we are currently investigating to achieve choreography adaptation that, beyond coordination, permits to face the challenges posed by the heterogeneity of FI services.

In this direction, we propose a way to enhance the previous CHOReOS approach in~\cite{FASE:13,serene:2013,Perucci:14,IEEE_SOFTWARE:2015}, and describes the preliminary steps we are undertaking within CHOReVOLUTION. The idea is to exploit Enterprise Integration Patterns~\cite{EIP:2004} (EIP) so as to deal with a form of choreography adaptation. Specifically, 
distributed coordination logic is first automatically synthesized out of the BPMN2 choreography specification. This logic is concrete service independent, meaning that it is synthesized by considering the expected interaction protocol specified for the roles instead of the one of the concrete services. This allows our approach to realize separation of concerns, hence possibly reusing the synthesized coordination logic when (late) binding different concrete services to the choreography roles. Then, 
service adapters are synthesized (only when needed) in order to correctly realize service-role binding. After a set of concrete participant services have been selected as suitable (yet not perfectly matching) candidates to play the choreography roles, adapters let the protocol of concrete services match the one specified for the played roles. Specifically, an adapter solves protocol mismatches between a concrete service and the played role by exploiting EIP as adaptation primitives and by suitably composing them to realize the required adaptation logic.

As a preliminary work, in this paper, we consider a subset of the Message Routing EIP hence addressing some specific protocol mismatches. A more complete treatment of a greater number of EIP, and related mismatches, is left for future work. 


%

The paper is structured as follow. Section~\ref{sec:settingthecontext} provides background notions on EIP. 
Section~\ref{sec:explanatoryexample} introduces an explanatory example. Then, Section~\ref{sec:choreosynthatadapt} describes how the synthesis process can be enhanced to deal with choreography adaptation through protocol coordination, protocol adaptation, and related complex data mappings. Section~\ref{sec:methodatwork} describes the proposed enhancement at work on the explanatory example. Related work is discussed in Section~\ref{sec:related}, and conclusions are given in Section~\ref{sec:conclusion}.

\section{Background notions on EIP}\label{sec:settingthecontext}

\vspace{-0.2cm}

This section 
provides basic notions about the subset of EIP that our approach exploits so far to deal with protocol adaptation issues.
As already introduced, to achieve adaptation, the protocol (i.e., operations signature plus interaction via messages exchange) of the concrete services may need to be adapted to the roles to be played in the BPMN2 choreography diagram given as input. This requires to implement a suitable notion of matching between protocols by means of \emph{complex data mappings} over both operation names and I/O messages. 

EIP offer more than one style for integrating heterogeneous (possibly mismatching) applications, i.e., \textit{File Transfer}, \textit{Shared Database}, \textit{Remote Procedure Invocation}, and \textit{Messaging}~\cite{EIP:2004}. We focus on the Messaging approach since we consider Web Services (WSs) as possible choreography participants, and WSs communicate through messages passing (e.g., request/response or one-way operation types). 


The Messaging approach uses the ``pipes-and-filters'' architectural style~\cite{SHAW_ARCH:1996} as base for ``streaming'' messages between endpoints. The latter (Filters) are connected with one another via Channels (Pipes). The producing endpoint sends messages to the channel, and the messages are retrieved by the consuming endpoint. There are different types of pipes and filters patterns, each of them dedicated to solve a particular integration aspect. Table~\ref{tbl:eips} describes the subset of EIP that we consider in this paper.

\newcolumntype{M}[1]{>{\centering\arraybackslash}m{#1}}
\begin{table}[ht]
\centering
\begin{tabular}{|M{2cm}|M{5cm}|M{7.5cm}|}
\hline
{\bf Name} & {\bf Description} & {\bf Figure} \\ \hline
Splitter & It splits a message into several ones, and sends the resulting messages to be processed independently &\includegraphics[scale=0.5]{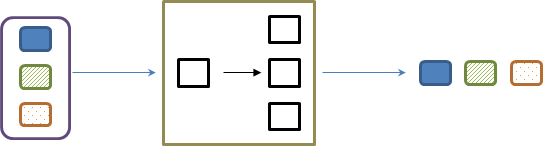}\\ \hline
Aggregator & It receives multiple messages and combines them into a single message. It is stateful and it must buffer the messages to be aggregated and determine when they are completed &\includegraphics[scale=0.5]{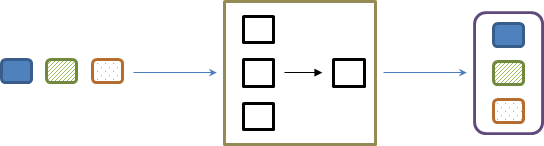}\\ \hline
Resequencer & It collects and re-orders messages and put them into an output channel in the specified order. Similarly to the Aggregator, it is stateful &\includegraphics[scale=0.5]{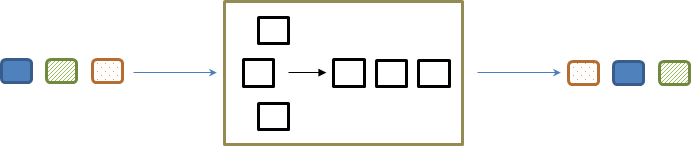}\\ \hline
Message Filter & It decides whether a message should be passed along or dropped based on some criteria respect to the header and/or content of the message &\includegraphics[scale=0.5]{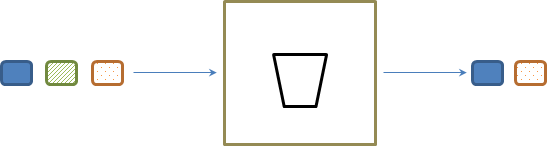}\\ \hline

\end{tabular}
\caption{Considered Message Routing EIP} \label{tbl:eips}
\end{table} 

The Aggregator, Splitter, Resequencer, and Message Filter belong to the class of Message Routing Patterns. This class of patterns is used to decouple a message source from the ultimate destination of the message following specific message routing policies as described in the table.



\section{Explanatory Example}\label{sec:explanatoryexample}

\vspace{-0.2cm}

The example introduced in this section is a very small portion of an \emph{In-store Marketing and Sale} choreography that was used by the EU CHOReOS project to demonstrate an \textit{Adaptive Customer Relationship Booster} system. The whole choreography was aimed at coordinating the activities of a client inside the shop in order to propose him/her tailored shopping offers and/or advertisements according to the user information (preferences, current shopping list, etc.) held by a	 shopping assistant application service.

\begin{figure}[h]
	\center
	\includegraphics[width=0.7\textwidth]{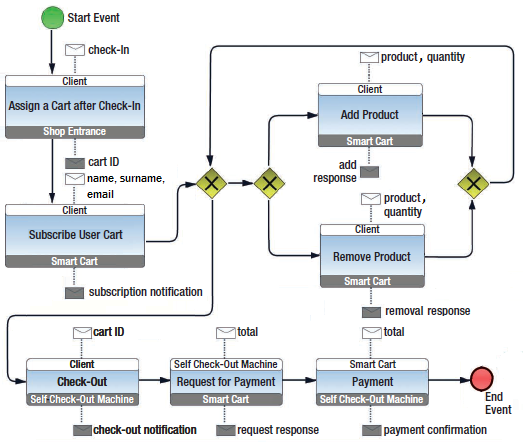}
	\caption{In-store Marketing and Sale choreography}
	\label{fig:MarketingAndSaleChor}
\end{figure}

Figure~\ref{fig:MarketingAndSaleChor} reports a simplified choreography diagram realized by using the Eclipse BPMN2 modeler plugin\footnote{\url{http://www.eclipse.org/bpmn2-modeler/}}. The diagram also shows the input and output messages of each choreography task. Within the Eclipse BPMN2 modeler, messages are specified by using the XML schema, which is the default language for specifying BPMN2 messages.

The choreography is triggered by the \texttt{Client} entering the shop. A \texttt{Shop Entrance} service (not shown in the figure) detects the presence of a specific \texttt{Client} inside the store and assigns him a virtual cart. Once subscribed to the cart, the \texttt{Client} can add and remove products to and from it. 
Once the \texttt{Client} finishes shopping, the \texttt{Smart Cart} service allows for executing the payment by interacting  with the a \texttt{Self Check-out Machine}.

\section{Method Description}\label{sec:choreosynthatadapt}

\vspace{-0.2cm}

The problem we want to solve with our synthesis method concerns the 
{\em automatic realizability enforcement} of service choreographies. It can be informally phrased as follows: given a choreography specification and a set of existing services, externally coordinate and adapt their interaction so to fulfill the collaboration prescribed by the choreography specification. Note that this problem is fundamentally different from the widely known {\em realizability check} (checks whether the choreography can be realized by implementing each participant so that it conforms to the played role) and {\em conformance check} (checks whether the set of services satisfies the choreography specification) problems~\cite{Bultan:2011,Sal08,pascal12,Gwen:12,Basu-Bultan-POPL:12,GwenPascalFASE:13,Bultan:2012,Bultan:2014}. 

\smallskip 
\noindent The method we propose distinguishes between {\em protocol coordination} and {\em protocol adaptation}. 

\begin{figure}[h]
	\center
	\includegraphics[width=0.8\textwidth]{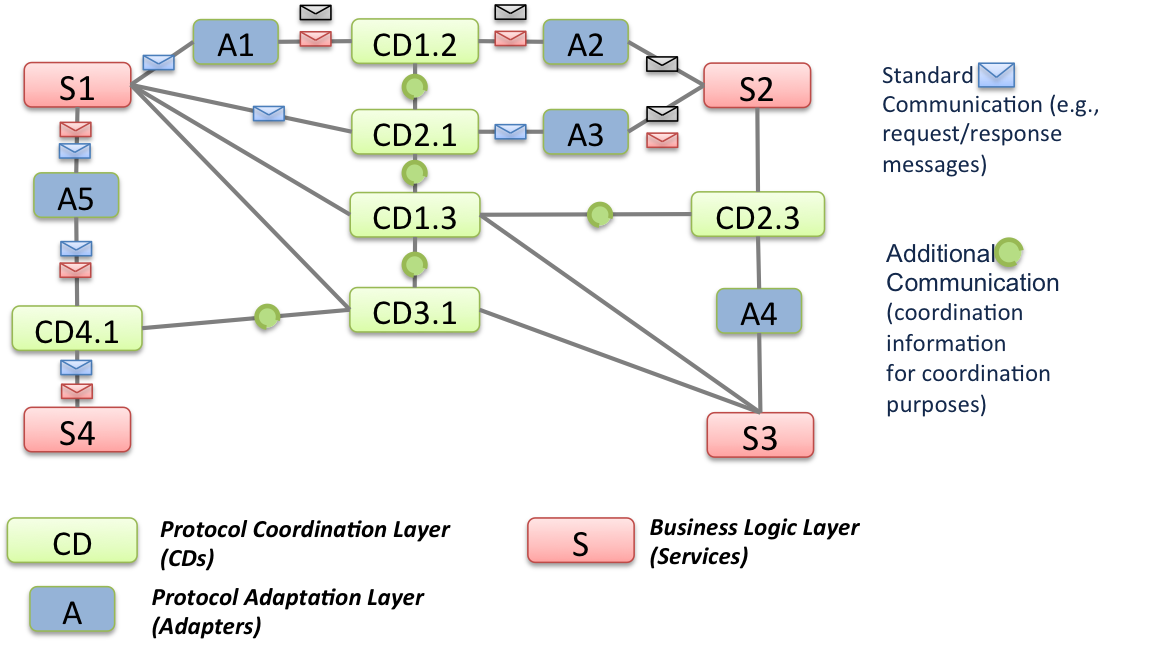}
	\caption{Architectural style with adapters}
	\label{fig:adapterArchStyle}
\end{figure}

Protocol coordination allows for preventing undesired interactions among (possibly adapted) services. That is, interactions not allowed by the choreography specification can happen when the services collaborate in an uncontrolled way. For instance, the \texttt{Client} is allowed to perform the \texttt{Add Product} task to add products to the \texttt{Smart Cart} (see the topmost right side of Figure~\ref{fig:MarketingAndSaleChor}). However, after paying and before the end event, a ``malicious'' \texttt{Client} may attempt at adding products to the cart, without paying for them. In order to achieve correct protocol coordination, additional software entities are synthesized and interposed among the participant services. They are hereafter referred to as {\em Coordination Delegates} (CDs). As described in~\cite{Foclasa14}, CDs proxify and coordinate the participant services' interaction. When interposed among the services, according to the architectural style shown in Figure~\ref{fig:adapterArchStyle}, CDs guarantee the collaboration specified by the choreography specification through distributed protocol coordination.

Protocol adaptation allows for dealing with services that do not exactly fit the choreography roles. That is, adapters are automatically synthesized to mediate the interaction Service-to-CD and CD-to-Service according to the choreography roles (see Figure~\ref{fig:adapterArchStyle}). Each Adapter is generated to bridge/mediate the concrete service protocol in order to exactly match the abstract participant protocol. In other words, adapters realize correct service-role binding by solving possible interoperability issues (e.g., protocol mismatches). 
By exploiting EIP, the generated adaptation logic is realized as a composition of instances of message routing patterns that realize complex I/O data mappings. For instance, adapters are able to map message data types, or reorder/merge/split the sequence of operation calls and/or related I/O messages.

Summing up, coordination and adaptation software entities are synthesized in order to: (i) proxify and coordinate the participant services' interaction to guarantee the global collaboration specified by the choreography (CDs), and (ii) adapt the concrete services' protocol to fit the protocol of the respective played roles (adapters). 

As already introduced, an important aspect here is that the coordination logic performed by the CDs is {\em service-independent} since it is based on the expected protocol of the participants as specified by the choreography, rather than on the protocol of the concrete services to be binded and coordinated. In this way {\em separation of concerns} is realized by separating pure coordination issues (i.e., undesired interactions) from adaptation/mediation ones (e.g., operation signature mismatches, data incompatibilities at the service interface level, and interaction mismatches). 

Section~\ref{implementation-overview} presents an overall and high-level description of the method implementation. Then, by focusing on the novel contribution of this paper with respect to our previous work done in CHOReOS, Section~\ref{adapters-generation} provides some details about the adapters generation step of our method.

\subsection{Overview of the Method Implementation} \label{implementation-overview}

\begin{figure}[h]
	\centering
	\includegraphics[width=0.7\textwidth]{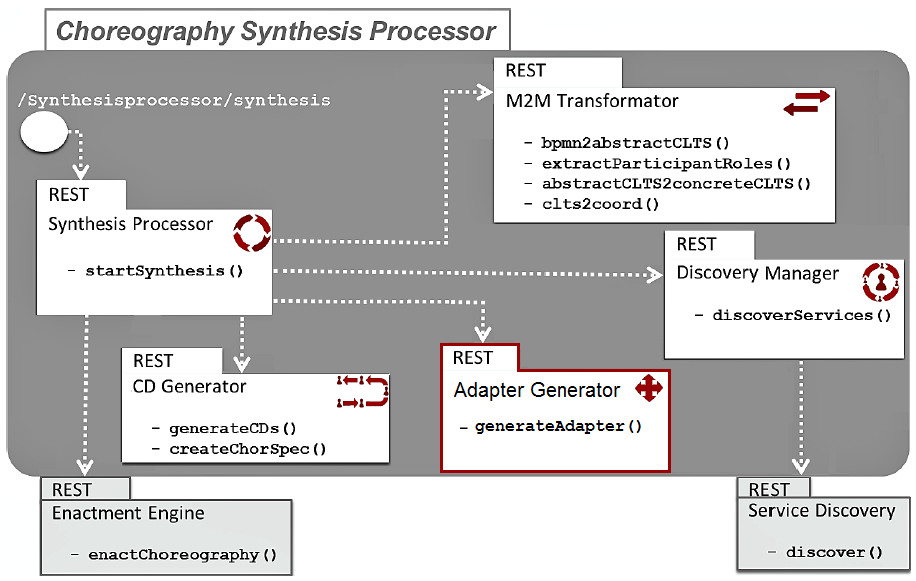}
	\caption{REST Architecture of the Synthesis Processor}
	\label{fig:restArchitecture}
\end{figure}

Our previous work within CHOReOS~\cite{FASE:13,serene:2013,Perucci:14,IEEE_SOFTWARE:2015}, and the related prototype implementation in the \texttt{CHOReOSynt} tool~\cite{Autili:14}, concerns the automated synthesis of CDs out of the choreography BPMN2 specification. In order to automatically synthesize adapters, this paper advances our previous work and proposes an extension of \texttt{CHOReOSynt} by introducing a new RESTful service called \texttt{Adapter Generator} (see Figure ~\ref{fig:restArchitecture}).

By referring to Figure~\ref{fig:restArchitecture}, the extension we propose allows for deriving both CDs and adapters starting from a BPMN 2.0 specification of the choreography, and a protocol specification of the concrete services selected as suitable participants. 
To this end, model transformations are employed and interoperation with the \texttt{Service Discovery} is required (not in the focus of this paper).
Both CDs and adapters, when deployed by the \texttt{Enactment Engine} (not in the focus of this paper), allow for enacting the choreography by realizing the distributed coordination and adaptation logic between the discovered services. 

%
%
%

The overall synthesis framework consists of the following RESTful services, and a set of  Eclipse plugins that have been developed to interact with such services.

\textbf{M2M Transformator --} The Model-to-Model (M2M) Transformator offers a set of transformations.

\textbf{CD Generator --} Starting from the choreography specification given as input (BPMN2 Choreography Diagram), the synthesis processor implements the approach formalized in~\cite{Foclasa14} in order to generate the needed CDs through the operation \texttt{generateCD()}.

\textbf{Synthesis Discovery Manager --} The Synthesis process and the Discovery process interact each other to retrieve, from the service base, the candidate services that are suitable for playing the participant roles required by the choreography specification. That is, the services whose (provided and required) operations and protocol are semantically compatible (possibly, not perfectly matching) with the operations and protocol of the generated CDs.

\textbf{Adapter Generator --} For each discovered service, if needed, an adapter is automatically synthesized through the operation \texttt{generateAdapter()}. The aim of the adapter is to bridge/mediate the protocol of a discovered service with the one of the played role as reified by the corresponding CD. 

%


\subsection{Focusing on Adapters Generation} \label{adapters-generation}

As already introduced, our method synthesizes adapters as WSs that implements a suitable composition of Message Routing EIP instances.

There are several frameworks and/or systems that implement/use EIP in order to integrate heterogeneous, and hence possibly mismatching, applications. We have chosen Spring Integration~\footnote{\url{http://projects.spring.io/spring-integration/}} because it implements most of the EIP, and it natively integrates with the Spring WSs project~\footnote{\url{http://projects.spring.io/spring-ws/}}.

\begin{figure}[h]
	\center
	\includegraphics[width=1\textwidth]{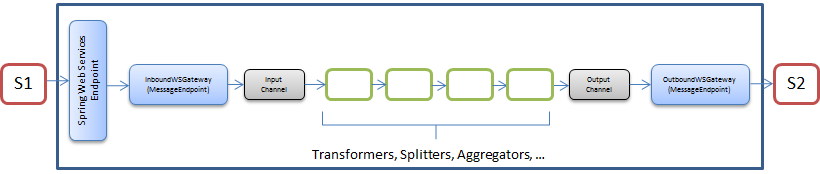}
	\caption{Adapter architecture}
	\label{fig:adapterArch}
\end{figure}

Figure~\ref{fig:adapterArch} describes the generic architecture of the generated adapters by using Spring WSs and Spring Integration. 
In particular, the Spring WSs \texttt{Endpoint} is the service that mediates the interaction between the service \texttt{S1} and the service \texttt{S2}. When \texttt{S1} invokes an operation \textit{op1} by sending a message \textit{m1}, the Endpoint receives the operation and put the message into the input channel by using \textit{Inbound WS Gateways}. A sequential composition (chain) of suitable EIP is then generated depending on the interoperability issues discovered (e.g., protocol mismatches) between the two interacting services, e.g., a concrete service and a CD in our setting. 
An handler of the chain can be one or more \textit{Message Routing EIP} as described in~\cite{EIP:2004} and in Table~\ref{tbl:eips}.

In order to support the automated identification of interoperability issues, our method exploits a slightly modified version of the Strawberry tool~\cite{Bertolino:2009}. In particular, Strawberry is used to automatically infer {\em data mappings} between different messages of two different WSs whose interaction needs to be mediated by an adapter. For instance, referring to our explanatory example, the two WSs could be the concrete \texttt{Client} service and the CD interposed between \texttt{Client} and \texttt{SmartCart}.

Strawberry is based on static data type analysis and testing. The former is used to analyze the type structure (e.g., the type structure of the XML Schema types of the messages in the WSDL of the considered WSs) of two messages belonging to the two different WSs. Based on the so far considered Message Routing EIP (Table~\ref{tbl:eips}), the adapters generation step reasons according to the following rules.

\begin{itemize}
\item {\bf Message splitting and aggregation}. Given two different messages of two different services $WS_1$ and $WS_2$ with types $t_1$ and $t_2$, respectively, the aim of this analysis is to check whether, e.g., $t_2$ is a subtype of $t_1$ (denoted as $t_1$$\preceq$$t_2$) according to the classical subtyping relation (i.e., signature inclusion): roughly, $t_1$ is contained in $t_2$. In our context, this means that either a message of type $t_2$ can be used to produce a message of type $t_1$ by means of message splitting or, vice versa, a message of type $t_1$ can be aggregated with other messages of $WS_1$ to produce a message of type $t_2$. Clearly, each of these other messages has to have a type $t$ such that $t$$\preceq$$t_2$. Whether it is a case of message splitting or message aggregation depends on the ``direction'' of the communication that one wants to achieve by realizing the inferred data mappings. For instance, let $t_2$ (resp., $t_1$) be the type of the input message of a required (resp., provided) operation, i.e., $WS_2$ behaves as the consumer and $WS_1$ as the provider. If $t_1$$\preceq$$t_2$ then $t_2$ has to be split to produce $t_1$. Otherwise, if $t_2$$\preceq$$t_1$, $t_2$ has to be aggregated with some other messages of $WS_2$ (for which $t_1$ is a subtype of their respective types) to produce $t_1$.

\item {\bf Message filtering}. It is worth to note that the data mappings inferred by Strawberry do not induce compositions of Splitter and Aggregator only. In fact, following the above discussion, for some pair of types $t_1$ and $t_2$, it can be the case that, e.g., no subtyping relation is defined for $t_2$ while $t_1$ has a subtyping relation with other message types of $WS_2$. This means that if $WS_2$ sends the message of type $t_2$ to $WS_1$, this message has to be consumed by exploiting a Message Filter.

\item {\bf Resequencing}. It can be the case that both $t_1$$\preceq$$t_2$ and $t_2$$\preceq$$t_1$ hold, denoted as $t_1$$\equiv$$t_2$. In other words, $t_1$ and $t_2$ are the same type. Continuing the above discussion, let us suppose that $WS_2$ sends a sequence of messages whose types are $t^1_2$$t^2_2$$\ldots$$t^n_2$ to $WS_1$ while the latter expects to receive a sequence of messages whose types are $t^1_1$$t^2_1$$\ldots$$t^n_1$. Suppose also that, for all $i$$\in$$\{$$1$$,$$\ldots$$,$$n$$\}$, Strawberry inferred that $t^i_1$$\equiv$$t^{\pi(i)}_2$ where $\pi$ is a permutation of $\{$$1$$,$$\ldots$$,$$n$$\}$, a Resequencer is needed in order to let $WS_2$ and $WS_1$ interact.
\end{itemize}

An important consideration here is that, in general, accounting for the messages' type structure only is not sufficient. In fact, one should check that two messages are also semantically correlated. The testing phase of Strawberry serves for this purpose. However, to perform testing with an acceptable accuracy, Strawberry requires the user to build an ad-hoc oracle. When testing cannot be performed, e.g., service providers do not offer a testing version of the services under analysis, the only thing that our method can do is to query the user (e.g., choreography designers, software architects) about confirming or not the semantic correlation of the data mappings inferred by Strawberry during the data type analysis. An alternative way to check message semantic correlation would be to exploit ontological information in a way similar to what is described in~\cite{Tivoli:2013}. For instance, the use of ontological information will be indeed useful in the future since we plan to extend our method so to account also for other kinds of EIP (e.g., Message Transformation EIP like Content Filter or Content Enricher~\cite{EIP:2004}).

\begin{figure}[h]
	\center
	\includegraphics[width=1\textwidth]{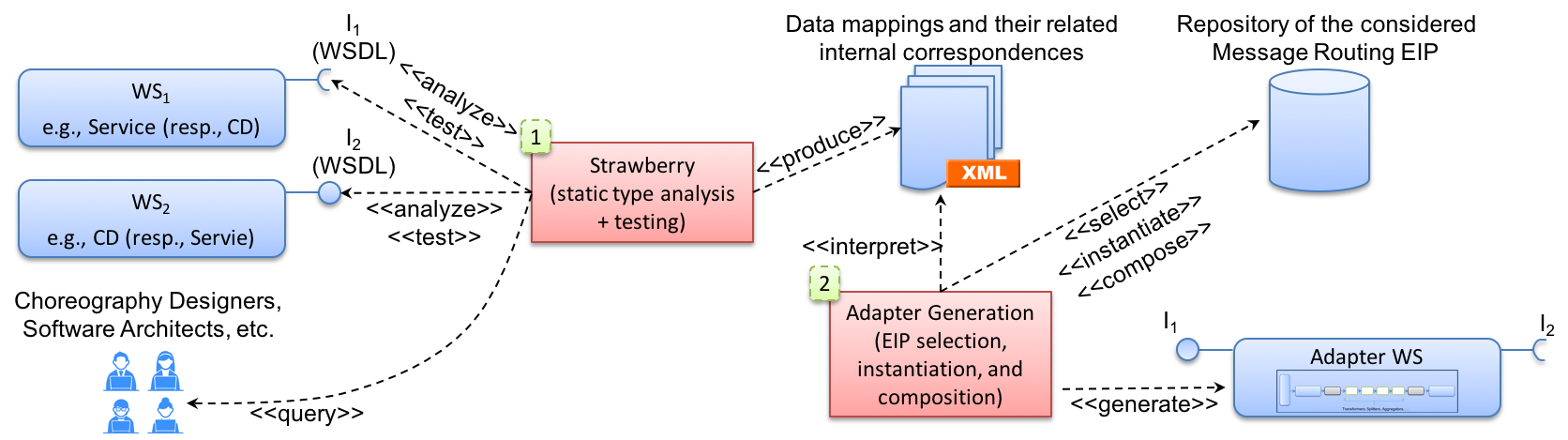}
	\caption{Adapters Generation Overview}
	\label{fig:adapterGen}
\end{figure}

Summing up, as shown in Figure~\ref{fig:adapterGen}, our adapters generation step can be overviewed as follows. Starting from (i) the specification of a consumer (resp., provider) participant service and (ii) the specification of the CD generated to coordinate the interaction between the consumer (resp., provider) role and the provider (resp., consumer) role, Strawberry produces a set of I/O data mappings and related internal data correspondences. As discussed above, these mappings allows for the detection of possible interoperability issues between the considered service and CD. These issues can arise when, e.g., the service required interface does not exactly match the CD provided one. This is done either completely automatically or semi-automatically by involving the user in the process. Then, an aptly coded model-to-code transformation is exploited in order to automatically generate the concrete adapter whose software architecture conforms to the one shown in Figure~\ref{fig:adapterArch}. This transformation interprets the elicited mappings and, according to the reasoning rules discussed above, select the required Message Routing EIP (among the considered ones), instantiate and compose them together to build the adapter.

\section{Method at Work}\label{sec:methodatwork}

\vspace{-0.2cm}

In this section we describe the proposed enhancement at work based on the explanatory example shown in Section~\ref{sec:explanatoryexample}. We focus on the \texttt{Add Product} Choreography Task. The Client role models an abstract consumer service that requires the operation \texttt{addProduct(product, quantity)} to add a specified \texttt{quantity} of \texttt{product} to the cart. The Smart Cart role models an abstract provider service representing the cart, hence providing the operation \texttt{addProduct(product, quantity)}. As already introduced, from the choreography specification in Figure~\ref{fig:MarketingAndSaleChor}, the coordination delegate \texttt{CD\pedice{Client\_SmartCart}} is synthesized. It coordinates (from outside) the interaction between the two services that will play the roles Client and Smart Cart respectively, in a way that the resulting collaboration realizes the specified choreography. Thus, \texttt{CD\pedice{Client\_SmartCart}} is a prosumer service (both a provider and a consumer of service operations) that provides, as well as requires, the operation \texttt{addProduct(product, quantity)}.

Now, let us suppose that a concrete consumer service \texttt{Client} has been selected to play the Client role. \texttt{Client} requires three operations, \texttt{addProduct(product)}, \texttt{setQuantity(quantity)} and \texttt{setPromotionCode(promotionCode)}, in order to perform the \texttt{Add Product} Choreography Task. That is, when adding a specified \texttt{quantity} of \texttt{product} to the cart it can also take advantage of some discount by further specifying a \texttt{promotionCode}. Now, let us also suppose that a concrete provider service \texttt{SmartCart} has been selected to play the Smart Cart role. \texttt{SmartCart} provides two operations, namely \texttt{addItem(item)} and \texttt{setAmount(amount)}, which allow its clients for adding a specified \texttt{amount} of \texttt{item} to the cart.

\subsection{Automated synthesis of data mappings} \label{sec:data-mappings}

\texttt{Client} requires an interface that does not match the one provided by \texttt{CD\pedice{Client\_SmartCart}}. Similarly, \texttt{SmartCart} provides an interface that does not match the one required by \texttt{CD\pedice{Client\_SmartCart}}. Thus, two adapters must be synthesized, one between \texttt{Client} and \texttt{CD\pedice{Client\_SmartCart}} and one between \texttt{CD\pedice{Client\_SmartCart}} and \texttt{SmartCart}.

Strawberry is able to infer the following data mappings. Note that no mapping has been retrieved for the message \texttt{Client.setPromotion Code.setPromotionCodeRequest} containing \texttt{promotionCode}.

{\scriptsize
\begin{itemize}
\item \texttt{Client.addProduct.addProductRequest} $\preceq$\\
\texttt{CD\pedice{Client\_SmartCart}.addProduct.addProductRequest}

\item \texttt{Client.setQuantity.setQuantityRequest} $\preceq$\\ \texttt{CD\pedice{Client\_SmartCart}.addProduct.addProductRequest}

\item \texttt{SmartCart.setAmount.setAmountRequest} $\preceq$\\ \texttt{CD\pedice{Client\_SmartCart}.addProduct.addProductRequest}

\item \texttt{SmartCart.addItem.addItemRequest} $\preceq$\\ \texttt{CD\pedice{Client\_SmartCart}.addProduct.addProductRequest}
\end{itemize}
}

Furthermore, the following internal data correspondences still concerning the inferred data mappings are also synthesized by Strawberry.

{\scriptsize
\begin{itemize}
\item \texttt{Client.addProduct.addProductRequest.product.id} \texttt{->}\\
\texttt{CD\pedice{Client\_SmartCart}.addProduct.addProductRequest.product.id}

\item \texttt{Client.addProduct.addProductRequest.product.description} \texttt{->}\\
\texttt{CD\pedice{Client\_SmartCart}.addProduct.addProductRequest.product.description}

\item \texttt{Client.setQuantity.setQuantityRequest.quantity} \texttt{->}\\ \texttt{CD\pedice{Client\_SmartCart}.addProduct.addProductRequest.quantity}

\item \texttt{CD\pedice{Client\_SmartCart}.addProduct.addProductRequest.quantity} \texttt{->}\\ \texttt{SmartCart.setAmount.setAmountRequest.amount}

\item  \texttt{CD\pedice{Client\_SmartCart}.addProduct.addProductRequest.product.id} \texttt{->}\\ \texttt{SmartCart.addItem.addItemRequest.itemCode} 

\item  \texttt{CD\pedice{Client\_SmartCart}.addProduct.addProductRequest.product.description} \texttt{->}\\ \texttt{SmartCart.addItem.addItemRequest.descr}
\end{itemize}
}

As described in Section~\ref{adapters-generation}, these data correspondences are crucial for supporting the automated synthesis of the required adapter. That is, the concrete routing logic that the EIP instances corresponding to the inferred data mappings, and composed to form the adapter, have to implement.

The following listings show the type structure of the messages in the above mappings (for the sake of simplicity \texttt{CD\pedice{Client\_SmartCart}} is referred to as CD).

{\scriptsize
\begin{lstlisting}[language=XML, caption=type structure of CD.addProduct.addProductRequest, label={lst:1}]
<xsd:schema version="1.0" targetNamespace="http://choreosynth.disim.univaq.it/">
   <xsd:complexType name="product">
     <xsd:sequence>
        <xsd:element name="id" type="xsd:string"></xsd:element>
        <xsd:element name="description" type="xsd:string"></xsd:element>
     </xsd:sequence>
   </xsd:complexType>
   <xsd:element name="quantity" type="xsd:int"></xsd:element>
</xsd:schema>
\end{lstlisting}
}

{\scriptsize
\begin{lstlisting}[language=XML, caption=type structure of Client.addProduct.addProductRequest, label={lst:2}]
<xsd:schema version="1.0" targetNamespace="http://choreosynth.disim.univaq.it/">
   <xsd:complexType name="product">
     <xsd:sequence>
        <xsd:element name="id" type="xsd:string"></xsd:element>
        <xsd:element name="description" type="xsd:string"></xsd:element>
     </xsd:sequence>
   </xsd:complexType>
</xsd:schema>
\end{lstlisting}
}

{\scriptsize
\begin{lstlisting}[language=XML, caption=type structure of Client.setQuantity.setQuantityRequest, label={lst:3}]
<xsd:schema version="1.0" targetNamespace="http://choreosynth.disim.univaq.it/">
   <xsd:element name="quantity" type="xsd:int"></xsd:element>
</xsd:schema>
\end{lstlisting}
}

{\scriptsize
\begin{lstlisting}[language=XML, caption=type structure of SmartCart.setAmount.setAmountRequest, label={lst:4}]
<xsd:schema version="1.0" targetNamespace="http://choreosynth.disim.univaq.it/">
   <xsd:element name="amount" type="xsd:int"></xsd:element>
</xsd:schema>
\end{lstlisting}
}

{\scriptsize
\begin{lstlisting}[language=XML, caption=type structure of SmartCart.addItem.addItemRequest, label={lst:5}]
<xsd:schema version="1.0" targetNamespace="http://choreosynth.disim.univaq.it/">
   <xsd:complexType name="item">
     <xsd:sequence>
        <xsd:element name="itemCode" type="xsd:string"></xsd:element>
        <xsd:element name="descr" type="xsd:string"></xsd:element>
     </xsd:sequence>
   </xsd:complexType>
</xsd:schema>
\end{lstlisting}
}

\subsection{Automated synthesis of adapters}
 
\begin{figure}[h]
	\center
	\includegraphics[width=1\textwidth]{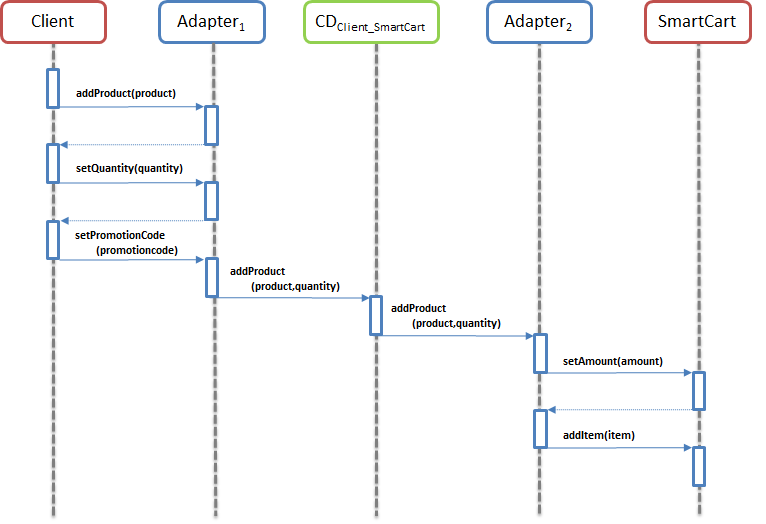}
	\caption{Adapter Example}
	\label{fig:adapterExample}
\end{figure}

As discussed in the previous section, our method exploits the data mappings inferred by Strawberry and the protocol specification of \texttt{Client}, \texttt{CD\pedice{Client\_SmartCart}}, and \texttt{SmartCart} in order to synthesize two adapters. They are \texttt{Adapter\pedice{1}} and \texttt{Adapter\pedice{2}} shown in Figure~\ref{fig:adapterExample}, which allow \texttt{Client} and \texttt{SmartCart} to communicate with \texttt{CD\pedice{Client\_SmartCart}} despite protocol mismatches.

\texttt{Adapter\pedice{1}} is generated by sequentially composing two EIP, namely Message Filter and Aggregator. The Message Filter takes as input from \texttt{Client} the three messages \texttt{addProductRequest}, \texttt{setQuantityRequest}, and \texttt{setPromotionCodeRequest}, and consumes \texttt{setPromotionCodeRequest} since there is no mapping inferred for it. This essentially means that the datum \texttt{promotionCode} is not needed to produce the data \texttt{product} and \texttt{quantity} contained in the \texttt{addProductRequest} message of \texttt{CD\pedice{Client\_SmartCart}}. Thus, at the end, Message Filter outputs \texttt{addProductRequest} and \texttt{setQuantityRequest}. Aggregator takes as input these two messages and merges them into one single message, hence producing the \texttt{addProductRequest} message of \texttt{CD\pedice{Client\_SmartCart}}.

\texttt{Adapter\pedice{2}} is generated by sequentially composing two EIP, namely Splitter and Resequencer. Splitter takes as input \texttt{addProductRequest} from \texttt{CD\pedice{Client\_SmartCart}} and splits it in the two separate messages containing the data {\tt product} and {\tt quantity}, respectively. Resequencer takes these two messages as input and reorders them hence invoking the operation \texttt{setAmount(amount)} of \texttt{SmartCart} and then the operation \texttt{addItem(item)} of the same service. Note that, thanks to the inferred data mappings, the message \texttt{setQuantityRequest} (resp., \texttt{addProductRequest}) contains the data required to produce the message \texttt{setAmountRequest} (resp., \texttt{addItemRequest}). The code generated for \texttt{Adapter\pedice{1}} and \texttt{Adapter\pedice{2}} is available at \url{http://choreos.disim.univaq.it}.

\section{Related}\label{sec:related}

\vspace{-0.2cm}

The mediation/adaptation of protocols have received attention since the early days of networking. Indeed, many efforts have been done in several directions including for example formal approaches to protocol conversion, like in~\cite{CalvertL90,Lam88}. Recently, with the emergence of web services and advocated universal interoperability, the research community has been studying solutions to the automatic mediation of business processes~\cite{VACULIN08}. However, most solutions are discussed informally, making it difficult to assess their respective advantages and drawbacks.

Spitznagel and Garlan propose an approach to formally specify adapter wrappers as protocol transformations, modularizing them, and reasoning about their properties, with the aim to resolve component mismatches~\cite{SPITZ_ICSE_03}. Although this formalizations supports modularization, automated synthesis is not treated, hence keeping the focus only on adapter design and specification.

Passerone et al. use a game theoretic approach for checking  whether incompatible component interfaces can be made compatible by inserting a converter between them which satisfies specified requirements. This approach is  able to automatically synthesize the converter~\cite{converter_synthesis}. In contrast to our method, their method needs as input a deadlock-free specification of the requirements that should be satisfied by the adapter, by delegating to the user the non-trivial task of specifying that.


In the context of {\em Reo connectors}, a number of related works are worth to be considered. The works described in~\cite{fmco08,tsc13} discuss different approaches to extract choreography specifications from BPMN and UML diagrams. 
Automated synthesis of choreography specifications from scenario-based specification is addressed in~\cite{scp11}. The main focus of the works described in~\cite{pdp14,esocc13} is on hybrid enforcement approaches. 

Rahm et al. propose a catalog of criteria for documenting the evaluations of schema matching systems~\cite{Erhard:2002}. In particular, the authors discuss various aspects that contribute to the match quality obtained as the result of an evaluation. In~\cite{Erhard:2002_COMA} the authors present a generic schema match system called COMA, which provides an extensible library of simple and hybrid match algorithms and supports a powerful framework for combining match results. This framework can be used for systematically evaluate different aspects of match processing, match direction, match candidate selection, and computation of combined similarity, and different matcher usages.

Paolucci et al. propose a base algorithm~\cite{Paolucci:2002} for semantic matching between service advertisements and service requests based on DAML-S, a DAML-based language for service description. The algorithm proposed differentiate between four degrees of matching and can be used for automatic dynamic discovery, selection and inter-operation of web services.

The approach described in~\cite{Gwen:2O12} enforces choreography realizability by automatically generating monitors. Each monitor acts as a local controller for its peer. This approach obtains monitors by iterating equivalence-checking steps between two centralized models of the whole system. The adopted monitor concept is similar to our CD. However, our approach synthesizes CDs without producing a centralized model of the whole system, hence preventing state explosion.
\section{Conclusion and Future Works}\label{sec:conclusion}

\vspace{-0.2cm}

In this paper, we propose a way to enhance the approach to the automatic synthesis of choreography-based systems we previously proposed in the context of the EU FP7 CHOReOS project. We report  the novel idea we are currently investigating within the EU H2020 CHOReVOLUTION project (a CHOReOS follow-up) to support choreography adaptation towards facing the challenges posed by the heterogeneity of Future Internet services. In particular, the idea is to automatically generate adapters by combining different EIP depending on a notion of I/O data mappings inference. 

In order to automatically synthesize adapters we propose an extension of our \texttt{CHOReOSynt} tool introducing a new RESTful service called \texttt{Adapter Generator} and we define the architecture of the genereted adapters by using Spring WSs and Spring Integration frameworks.
Referring to an explanatory example, we have shown two types of adaptation using some Message Routing Patterns, namely Message Filter, Aggregator, Splitter, and Resequencer. 

The relevance of exploiting EIP to build adapters resides in the fact that the adapters generated by our method have a modular architecture since they are structured as a composition of independent EIP, each of them corresponding to the solution of a recurring protocol mismatch. This represents an overall advantage of our approach with respect to the work in the state of the art (Section~\ref{sec:related}). Because of the way adapters are modularly structured as a composition of fundamentals mediation primitives, the automatic generation of their actual code is viable and can be achieved with little effort.

As future work, in order to achieve the even more ambitious objectives of the CHOReVOLUTION project and to improve the applicability of the approach, we plan to extend it to deal with security issues concerning scenarios in which multiple services involved in a choreography belong to different security domains governed by different authorities and use different identity attributes that are utilized in their access control polices. This should be achieved by integrating EIPs with Security Patterns~\cite{SecurityPatters:2005}. 

Furthermore, we plan to investigate how to support the automated synthesis of more complex adapters realized by combining other classes of EIP, e.g., Message Transformation Patterns such as Content Enricher, Content Filter, and Transformer. In general, this class of patterns provide a solution for achieving (semantic) interoperability between applications that communicate via message passing but rarely agree on a common data format. Message Transformation Patterns offer a general solution to possible differences in the data format of the exchanged messages. Thus, the ability to deal with these patterns in a systematic way would enable a finer form of adaptation concerning mismatches at the level of the semantics of the exchanged messages.

\section*{Acknowledgment}

This research work has been supported by the Ministry of Education, Universities and Research, prot. 2012E47TM2 (project IDEAS - Integrated Design and Evolution of Adaptive Systems), by the European Union's H2020 Programme under grant agreement number 644178 (project CHOReVOLUTION - Automated Synthesis of Dynamic and Secured Choreographies for the Future Internet), and by the Ministry of Economy and Finance, Cipe resolution n. 135/2012 (project INCIPICT - INnovating CIty Planning through Information and Communication Technologies).

\bibliography{biblio}

\begin{thebibliography}{10}
\providecommand{\bibitemdeclare}[2]{}
\providecommand{\surnamestart}{}
\providecommand{\surnameend}{}
\providecommand{\urlprefix}{Available at }
\providecommand{\url}[1]{\texttt{#1}}
\providecommand{\href}[2]{\texttt{#2}}
\providecommand{\urlalt}[2]{\href{#1}{#2}}
\providecommand{\doi}[1]{doi:\urlalt{http://dx.doi.org/#1}{#1}}
\providecommand{\bibinfo}[2]{#2}

\bibitemdeclare{article}{IEEE_SOFTWARE:2015}
\bibitem{IEEE_SOFTWARE:2015}
\bibinfo{author}{M.~\surnamestart Autili\surnameend},
  \bibinfo{author}{P.~\surnamestart Inverardi\surnameend} \&
  \bibinfo{author}{M.~\surnamestart Tivoli\surnameend} (\bibinfo{year}{2015}):
  \emph{\bibinfo{title}{Automated Synthesis of Service Choreographies}}.
\newblock {\sl \bibinfo{journal}{Software, IEEE}}
  \bibinfo{volume}{32}(\bibinfo{number}{1}), pp. \bibinfo{pages}{50--57},
  \doi{10.1109/MS.2014.131}.

\bibitemdeclare{incollection}{serene:2013}
\bibitem{serene:2013}
\bibinfo{author}{Marco \surnamestart Autili\surnameend},
  \bibinfo{author}{Amleto \surnamestart Di~Salle\surnameend} \&
  \bibinfo{author}{Massimo \surnamestart Tivoli\surnameend}
  (\bibinfo{year}{2013}): \emph{\bibinfo{title}{Synthesis of Resilient
  Choreographies}}.
\newblock In: {\sl \bibinfo{booktitle}{SERENE}}, \bibinfo{series}{LNCS 8166},
  \doi{10.1007/978-3-642-40894-6\_8}.

\bibitemdeclare{incollection}{FASE:13}
\bibitem{FASE:13}
\bibinfo{author}{Marco \surnamestart Autili\surnameend},
  \bibinfo{author}{Davide \surnamestart Ruscio\surnameend},
  \bibinfo{author}{Amleto \surnamestart Di~Salle\surnameend},
  \bibinfo{author}{Paola \surnamestart Inverardi\surnameend} \&
  \bibinfo{author}{Massimo \surnamestart Tivoli\surnameend}
  (\bibinfo{year}{2013}): \emph{\bibinfo{title}{A Model-Based Synthesis Process
  for Choreography Realizability Enforcement}}.
\newblock In: {\sl \bibinfo{booktitle}{FASE}}, {\sl \bibinfo{series}{LNCS}}
  \bibinfo{volume}{7793}, \doi{10.1007/978-3-642-37057-1\_4}.

\bibitemdeclare{inproceedings}{Autili:14}
\bibitem{Autili:14}
\bibinfo{author}{Marco \surnamestart Autili\surnameend},
  \bibinfo{author}{Davide~Di \surnamestart Ruscio\surnameend},
  \bibinfo{author}{Amleto~Di \surnamestart Salle\surnameend} \&
  \bibinfo{author}{Alexander \surnamestart Perucci\surnameend}
  (\bibinfo{year}{2014}): \emph{\bibinfo{title}{CHOReOSynt: enforcing
  choreography realizability in the future internet}}.
\newblock In: {\sl \bibinfo{booktitle}{Proceedings of the 22nd {ACM} {SIGSOFT}
  International Symposium on Foundations of Software Engineering, (FSE-22),
  Hong Kong, China, November 16 - 22, 2014}}, pp. \bibinfo{pages}{723--726},
  \doi{10.1145/2635868.2661667}.

\bibitemdeclare{inproceedings}{Foclasa14}
\bibitem{Foclasa14}
\bibinfo{author}{Marco \surnamestart Autili\surnameend} \&
  \bibinfo{author}{Massimo \surnamestart Tivoli\surnameend}
  (\bibinfo{year}{2015}): \emph{\bibinfo{title}{Distributed Enforcement of
  Service Choreographies}}.
\newblock In: {\sl \bibinfo{booktitle}{Proceedings 13th International Workshop
  on Foundations of Coordination Languages and Self-Adaptive Systems, {FOCLASA}
  2014, Rome, Italy, 6th September 2014.}}, pp. \bibinfo{pages}{18--35},
  \doi{10.4204/EPTCS.175.2}.

\bibitemdeclare{inproceedings}{Bultan:2011}
\bibitem{Bultan:2011}
\bibinfo{author}{Samik \surnamestart Basu\surnameend} \&
  \bibinfo{author}{Tevfik \surnamestart Bultan\surnameend}
  (\bibinfo{year}{2011}): \emph{\bibinfo{title}{Choreography conformance via
  synchronizability}}.
\newblock In: {\sl \bibinfo{booktitle}{Proc. of WWW '11}},
  \doi{10.1145/1963405.1963516}.

\bibitemdeclare{inproceedings}{Bultan:2014}
\bibitem{Bultan:2014}
\bibinfo{author}{Samik \surnamestart Basu\surnameend} \&
  \bibinfo{author}{Tevfik \surnamestart Bultan\surnameend}
  (\bibinfo{year}{2014}): \emph{\bibinfo{title}{Automatic verification of
  interactions in asynchronous systems with unbounded buffers}}.
\newblock In: {\sl \bibinfo{booktitle}{{ACM/IEEE} International Conference on
  Automated Software Engineering, {ASE} '14, Vasteras, Sweden - September 15 -
  19, 2014}}, pp. \bibinfo{pages}{743--754}, \doi{10.1145/2642937.2643016}.

\bibitemdeclare{inproceedings}{Basu-Bultan-POPL:12}
\bibitem{Basu-Bultan-POPL:12}
\bibinfo{author}{Samik \surnamestart Basu\surnameend}, \bibinfo{author}{Tevfik
  \surnamestart Bultan\surnameend} \& \bibinfo{author}{Meriem \surnamestart
  Ouederni\surnameend} (\bibinfo{year}{2012}): \emph{\bibinfo{title}{Deciding
  choreography realizability}}.
\newblock \bibinfo{series}{POPL}, \bibinfo{publisher}{ACM},
  \doi{10.1145/2103656.2103680}.

\bibitemdeclare{inproceedings}{Bertolino:2009}
\bibitem{Bertolino:2009}
\bibinfo{author}{Antonia \surnamestart Bertolino\surnameend},
  \bibinfo{author}{Paola \surnamestart Inverardi\surnameend},
  \bibinfo{author}{Patrizio \surnamestart Pelliccione\surnameend} \&
  \bibinfo{author}{Massimo \surnamestart Tivoli\surnameend}
  (\bibinfo{year}{2009}): \emph{\bibinfo{title}{Automatic Synthesis of Behavior
  Protocols for Composable Web-services}}.
\newblock In: {\sl \bibinfo{booktitle}{Proceedings of the the 7th Joint Meeting
  of the European Software Engineering Conference and the ACM SIGSOFT Symposium
  on The Foundations of Software Engineering}}, \bibinfo{series}{ESEC/FSE '09},
  pp. \bibinfo{pages}{141--150}, \doi{10.1145/1595696.1595719}.

\bibitemdeclare{article}{CalvertL90}
\bibitem{CalvertL90}
\bibinfo{author}{Kenneth~L. \surnamestart Calvert\surnameend} \&
  \bibinfo{author}{Simon~S. \surnamestart Lam\surnameend}
  (\bibinfo{year}{1990}): \emph{\bibinfo{title}{Formal Methods for Protocol
  Conversion}}.
\newblock {\sl \bibinfo{journal}{IEEE Journal on Selected Areas in
  Communications}} \bibinfo{volume}{8}(\bibinfo{number}{1}),
  \doi{10.1109/49.46852}.

\bibitemdeclare{inproceedings}{Erhard:2002}
\bibitem{Erhard:2002}
\bibinfo{author}{Hong~Hai \surnamestart Do\surnameend}, \bibinfo{author}{Sergey
  \surnamestart Melnik\surnameend} \& \bibinfo{author}{Erhard \surnamestart
  Rahm\surnameend} (\bibinfo{year}{2002}): \emph{\bibinfo{title}{Comparison of
  Schema Matching Evaluations}}.
\newblock In: {\sl \bibinfo{booktitle}{Web, Web-Services, and Database Systems,
  NODe 2002 Web and Database-Related Workshops, Erfurt, Germany, October 7-10,
  2002, Revised Papers}}, pp. \bibinfo{pages}{221--237},
  \doi{10.1007/3-540-36560-5\_17}.

\bibitemdeclare{inproceedings}{Erhard:2002_COMA}
\bibitem{Erhard:2002_COMA}
\bibinfo{author}{Hong~Hai \surnamestart Do\surnameend} \&
  \bibinfo{author}{Erhard \surnamestart Rahm\surnameend}
  (\bibinfo{year}{2002}): \emph{\bibinfo{title}{{COMA} - {A} System for
  Flexible Combination of Schema Matching Approaches}}.
\newblock In: {\sl \bibinfo{booktitle}{{VLDB} 2002, Proceedings of 28th
  International Conference on Very Large Data Bases, August 20-23, 2002, Hong
  Kong, China}}, pp. \bibinfo{pages}{610--621}, \doi{10.1016/1287369.1287422}.

\bibitemdeclare{misc}{FIRE:15}
\bibitem{FIRE:15}
\bibinfo{author}{\surnamestart {European Commission}\surnameend}
  (\bibinfo{year}{2015}): \emph{\bibinfo{title}{{Digital Agenda for Europe -
  Future Internet Research and Experimentation (FIRE) initiative}}}.
\newblock
  \urlprefix\url{https://ec.europa.eu/digital-agenda/en/future-internet-research-and-experimentation}.

\bibitemdeclare{incollection}{Gwen:12}
\bibitem{Gwen:12}
\bibinfo{author}{Gregor \surnamestart G{\"o}ssler\surnameend} \&
  \bibinfo{author}{Gwen \surnamestart Sala{\"u}n\surnameend}
  (\bibinfo{year}{2012}): \emph{\bibinfo{title}{Realizability of Choreographies
  for Services Interacting Asynchronously}}.
\newblock In: {\sl \bibinfo{booktitle}{FACS}}, {\sl \bibinfo{series}{LNCS}}
  \bibinfo{volume}{7253}, pp. \bibinfo{pages}{151--167},
  \doi{10.1007/978-3-642-35743-5\_10}.

\bibitemdeclare{incollection}{GwenPascalFASE:13}
\bibitem{GwenPascalFASE:13}
\bibinfo{author}{Matthias \surnamestart G{\"u}demann\surnameend},
  \bibinfo{author}{Pascal \surnamestart Poizat\surnameend},
  \bibinfo{author}{Gwen \surnamestart Sala{\"u}n\surnameend} \&
  \bibinfo{author}{Alexandre \surnamestart Dumont\surnameend}
  (\bibinfo{year}{2013}): \emph{\bibinfo{title}{VerChor: A Framework for
  Verifying Choreographies}}.
\newblock In: {\sl \bibinfo{booktitle}{FASE}}, {\sl \bibinfo{series}{LNCS}}
  \bibinfo{volume}{7793}, pp. \bibinfo{pages}{226--230},
  \doi{10.1007/978-3-642-37057-1\_16}.

\bibitemdeclare{inproceedings}{Gwen:2O12}
\bibitem{Gwen:2O12}
\bibinfo{author}{Matthias \surnamestart G{\"{u}}demann\surnameend},
  \bibinfo{author}{Gwen \surnamestart Sala{\"{u}}n\surnameend} \&
  \bibinfo{author}{Meriem \surnamestart Ouederni\surnameend}
  (\bibinfo{year}{2012}): \emph{\bibinfo{title}{Counterexample Guided Synthesis
  of Monitors for Realizability Enforcement}}.
\newblock In: {\sl \bibinfo{booktitle}{Automated Technology for Verification
  and Analysis - 10th International Symposium, {ATVA} 2012, Thiruvananthapuram,
  India, October 3-6, 2012. Proceedings}}, pp. \bibinfo{pages}{238--253},
  \doi{10.1007/978-3-642-33386-6\_20}.

\bibitemdeclare{book}{EIP:2004}
\bibitem{EIP:2004}
\bibinfo{author}{Gregor \surnamestart Hohpe\surnameend} \&
  \bibinfo{author}{Bobby \surnamestart Woolf\surnameend}
  (\bibinfo{year}{2004}): \emph{\bibinfo{title}{Enterprise Integration
  Patterns: Designing, Building, and Deploying Messaging Solutions - Fiftheenth
  printing 2011}}.
\newblock \bibinfo{publisher}{Addison-Wesley Longman Publishing Co., Inc.},
  \bibinfo{address}{Boston, MA, USA}.

\bibitemdeclare{inproceedings}{Tivoli:2013}
\bibitem{Tivoli:2013}
\bibinfo{author}{Paola \surnamestart Inverardi\surnameend} \&
  \bibinfo{author}{Massimo \surnamestart Tivoli\surnameend}
  (\bibinfo{year}{2013}): \emph{\bibinfo{title}{Automatic Synthesis of Modular
  Connectors via Composition of Protocol Mediation Patterns}}.
\newblock In: {\sl \bibinfo{booktitle}{Proceedings of ICSE'13}},
  \doi{10.1109/ICSE.2013.6606546}.

\bibitemdeclare{inproceedings}{esocc13}
\bibitem{esocc13}
\bibinfo{author}{Sung-Shik T.~Q. \surnamestart Jongmans\surnameend} \&
  \bibinfo{author}{Farhad \surnamestart Arbab\surnameend}
  (\bibinfo{year}{2013}): \emph{\bibinfo{title}{Global Consensus through Local
  Synchronization}}.
\newblock In: {\sl \bibinfo{booktitle}{ESOCC Workshops}}, pp.
  \bibinfo{pages}{174--188}, \doi{10.1007/978-3-642-45364-9\_15}.

\bibitemdeclare{inproceedings}{pdp14}
\bibitem{pdp14}
\bibinfo{author}{Sung-Shik T.~Q. \surnamestart Jongmans\surnameend},
  \bibinfo{author}{Francesco \surnamestart Santini\surnameend} \&
  \bibinfo{author}{Farhad \surnamestart Arbab\surnameend}
  (\bibinfo{year}{2014}): \emph{\bibinfo{title}{Partially-Distributed
  Coordination with Reo}}.
\newblock In: {\sl \bibinfo{booktitle}{PDP}}, pp. \bibinfo{pages}{697--706},
  \doi{10.1109/PDP.2014.19}.

\bibitemdeclare{inproceedings}{fmco08}
\bibitem{fmco08}
\bibinfo{author}{Natallia \surnamestart Kokash\surnameend} \&
  \bibinfo{author}{Farhad \surnamestart Arbab\surnameend}
  (\bibinfo{year}{2008}): \emph{\bibinfo{title}{Formal Behavioral Modeling and
  Compliance Analysis for Service-Oriented Systems}}.
\newblock In: {\sl \bibinfo{booktitle}{FMCO}}, pp. \bibinfo{pages}{21--41},
  \doi{10.1007/978-3-642-04167-9\_2}.

\bibitemdeclare{article}{tsc13}
\bibitem{tsc13}
\bibinfo{author}{Natallia \surnamestart Kokash\surnameend} \&
  \bibinfo{author}{Farhad \surnamestart Arbab\surnameend}
  (\bibinfo{year}{2013}): \emph{\bibinfo{title}{Formal Design and Verification
  of Long-Running Transactions with Extensible Coordination Tools}}.
\newblock {\sl \bibinfo{journal}{IEEE T. Services Computing}}
  \bibinfo{volume}{6}(\bibinfo{number}{2}), pp. \bibinfo{pages}{186--200},
  \doi{10.1109/TSC.2011.46}.

\bibitemdeclare{article}{Lam88}
\bibitem{Lam88}
\bibinfo{author}{Simon~S. \surnamestart Lam\surnameend} (\bibinfo{year}{1988}):
  \emph{\bibinfo{title}{Correction to "Protocol Conversion"}}.
\newblock {\sl \bibinfo{journal}{IEEE Trans. Software Eng.}}
  \bibinfo{volume}{14}(\bibinfo{number}{9}), \doi{10.1109/32.6181}.

\bibitemdeclare{article}{scp11}
\bibitem{scp11}
\bibinfo{author}{Sun \surnamestart Meng\surnameend}, \bibinfo{author}{Farhad
  \surnamestart Arbab\surnameend} \& \bibinfo{author}{Christel \surnamestart
  Baier\surnameend} (\bibinfo{year}{2011}): \emph{\bibinfo{title}{Synthesis of
  Reo circuits from scenario-based interaction specifications}}.
\newblock {\sl \bibinfo{journal}{Sci. Comput. Program.}}
  \bibinfo{volume}{76}(\bibinfo{number}{8}), pp. \bibinfo{pages}{651--680},
  \doi{10.1007/978-3-540-78743-3\_12}.

\bibitemdeclare{inproceedings}{Paolucci:2002}
\bibitem{Paolucci:2002}
\bibinfo{author}{Massimo \surnamestart Paolucci\surnameend},
  \bibinfo{author}{Takahiro \surnamestart Kawamura\surnameend},
  \bibinfo{author}{Terry~R. \surnamestart Payne\surnameend} \&
  \bibinfo{author}{Katia~P. \surnamestart Sycara\surnameend}
  (\bibinfo{year}{2002}): \emph{\bibinfo{title}{Semantic Matching of Web
  Services Capabilities}}.
\newblock In: {\sl \bibinfo{booktitle}{The Semantic Web - {ISWC} 2002, First
  International Semantic Web Conference, Sardinia, Italy, June 9-12, 2002,
  Proceedings}}, pp. \bibinfo{pages}{333--347},
  \doi{10.1007/3-540-48005-6\_26}.

\bibitemdeclare{inproceedings}{converter_synthesis}
\bibitem{converter_synthesis}
\bibinfo{author}{Roberto \surnamestart Passerone\surnameend},
  \bibinfo{author}{Luca~De \surnamestart Alfaro\surnameend},
  \bibinfo{author}{Thomas~A. \surnamestart Henzinger\surnameend} \&
  \bibinfo{author}{Alberto~L. \surnamestart Sangiovanni-Vincentelli\surnameend}
  (\bibinfo{year}{2002}): \emph{\bibinfo{title}{Convertibility Verification and
  Converter Synthesis: Two Faces of the Same Coin}}.
\newblock In: {\sl \bibinfo{booktitle}{ICCAD}}, \doi{10.1145/774572.774592}.

\bibitemdeclare{inproceedings}{pascal12}
\bibitem{pascal12}
\bibinfo{author}{Pascal \surnamestart Poizat\surnameend} \&
  \bibinfo{author}{Gwen \surnamestart Sala{\"u}n\surnameend}
  (\bibinfo{year}{2012}): \emph{\bibinfo{title}{{Checking the Realizability of
  BPMN 2.0 Choreographies}}}.
\newblock In: {\sl \bibinfo{booktitle}{{Proc. of SAC 2012}}},
  \doi{10.1145/2245276.2232095}.

\bibitemdeclare{inproceedings}{Sal08}
\bibitem{Sal08}
\bibinfo{author}{Gwen \surnamestart Sala\"{u}n\surnameend}
  (\bibinfo{year}{2008}): \emph{\bibinfo{title}{Generation of Service Wrapper
  Protocols from Choreography Specifications}}.
\newblock In: {\sl \bibinfo{booktitle}{Proc. of SEFM}},
  \doi{10.1109/SEFM.2008.42}.

\bibitemdeclare{article}{Bultan:2012}
\bibitem{Bultan:2012}
\bibinfo{author}{Gwen \surnamestart Sala{\"{u}}n\surnameend},
  \bibinfo{author}{Tevfik \surnamestart Bultan\surnameend} \&
  \bibinfo{author}{Nima \surnamestart Roohi\surnameend} (\bibinfo{year}{2012}):
  \emph{\bibinfo{title}{Realizability of Choreographies Using Process Algebra
  Encodings}}.
\newblock {\sl \bibinfo{journal}{{IEEE} T. Services Computing}}
  \bibinfo{volume}{5}(\bibinfo{number}{3}), pp. \bibinfo{pages}{290--304},
  \doi{10.1109/TSC.2011.9}.

\bibitemdeclare{inproceedings}{Perucci:14}
\bibitem{Perucci:14}
\bibinfo{author}{Amleto~Di \surnamestart Salle\surnameend},
  \bibinfo{author}{Paola \surnamestart Inverardi\surnameend} \&
  \bibinfo{author}{Alexander \surnamestart Perucci\surnameend}
  (\bibinfo{year}{2014}): \emph{\bibinfo{title}{Towards Adaptable and Evolving
  Service Choreography in the Future Internet}}.
\newblock In: {\sl \bibinfo{booktitle}{2014 {IEEE} World Congress on Services,
  {SERVICES} 2014, Anchorage, AK, USA, June 27 - July 2, 2014}}, pp.
  \bibinfo{pages}{333--337}, \doi{10.1109/SERVICES.2014.65}.

\bibitemdeclare{book}{SecurityPatters:2005}
\bibitem{SecurityPatters:2005}
\bibinfo{author}{Markus \surnamestart Schumacher\surnameend},
  \bibinfo{author}{Eduardo \surnamestart Fernandez-Buglioni\surnameend},
  \bibinfo{author}{Duane \surnamestart Hybertson\surnameend},
  \bibinfo{author}{Frank \surnamestart Buschmann\surnameend} \&
  \bibinfo{author}{Peter \surnamestart Sommerlad\surnameend}
  (\bibinfo{year}{2005}): \emph{\bibinfo{title}{Security Patterns Integrating
  Security and Systems Engineering}}.
\newblock \bibinfo{publisher}{John Wiley and Sons Ltd}.

\bibitemdeclare{book}{SHAW_ARCH:1996}
\bibitem{SHAW_ARCH:1996}
\bibinfo{author}{Mary \surnamestart Shaw\surnameend} \& \bibinfo{author}{David
  \surnamestart Garlan\surnameend} (\bibinfo{year}{1996}):
  \emph{\bibinfo{title}{Software architecture - perspectives on an emerging
  discipline}}.
\newblock \bibinfo{publisher}{Prentice Hall}.

\bibitemdeclare{inproceedings}{SPITZ_ICSE_03}
\bibitem{SPITZ_ICSE_03}
\bibinfo{author}{Bridget \surnamestart Spitznagel\surnameend} \&
  \bibinfo{author}{David \surnamestart Garlan\surnameend}
  (\bibinfo{year}{2003}): \emph{\bibinfo{title}{A Compositional Formalization
  of Connector Wrappers}}.
\newblock In: {\sl \bibinfo{booktitle}{ICSE}}, \doi{10.1109/ICSE.2003.1201216}.

\bibitemdeclare{inproceedings}{VACULIN08}
\bibitem{VACULIN08}
\bibinfo{author}{Roman \surnamestart Vacul\'in\surnameend},
  \bibinfo{author}{Roman \surnamestart Neruda\surnameend} \&
  \bibinfo{author}{Katia~P. \surnamestart Sycara\surnameend}
  (\bibinfo{year}{2008}): \emph{\bibinfo{title}{An Agent for Asymmetric Process
  Mediation in Open Environments.}}
\newblock In: {\sl \bibinfo{booktitle}{SOCASE}},
  \doi{10.1007/978-3-540-79968-9\_9}.

\end{thebibliography}

\bibliographystyle{eptcs}

\end{document}